\documentclass[12pt]{article}
\input{epsf}
\usepackage{amssymb,amsmath,latexsym,epsf}
\title{A Theory for Quantum Accelerator Modes in Atom Optics}
\vspace{0.5cm}

\author{Shmuel Fishman$^{1}$, Italo Guarneri$^{2,3,4}$, 
Laura Rebuzzini$^{2}$\\
{\small $^1$ Physics Department, Technion, Haifa 32000, Israel}\\
{\small $^2$ Centro di Ricerca per i Sistemi Dinamici}\\ 
{\small Universit\`a dell'Insubria a Como, via Valleggio 11, 22100 Como, Italy}\\
{\small $^3$  Istituto Nazionale per la Fisica della Materia, 
via Celoria 16, 20133 Milano, Italy}\\
{\small $^4$ Istituto Nazionale di Fisica Nucleare, Sezione di Pavia, 
via Bassi 6, 27100 Pavia, Italy}} 
\vspace{.2cm}

\newcommand{\Uop}{{\hat U}}
\newcommand{\Kop}{{\hat K}}
\newcommand{\Rop}{{\hat R}}
\newcommand{\Sop}{{\hat S}}

\newcommand{\En} {{\mathcal E}}
\newcommand{\kcl}{{\tilde k}}
\newcommand{\ti}{\mathtt{t}}
\newcommand{\jt}{{\cal J}}
\newcommand{\pG}{{\mathfrak p}}
\newcommand{\jG}{{\mathfrak j}}
\newcommand{\Up}{{\hat{\mathcal U}}}
\newcommand{\Kp}{{\hat{\mathcal K}}}
\newcommand{\Rp}{{\hat{\mathcal R}}}
\newcommand{\Hp}{{\hat{\mathcal H}}}
\newcommand{\Np}{{\hat{\mathcal N}}}
\newcommand{\Mi}{\mbox{\rm\small min}}
\newcommand{\Mx}{\mbox{\rm\small max}}
\newcommand{\Dp}{\partial}

\begin{document}

\maketitle

\begin{abstract}
{ Unexpected accelerator modes were recently observed experimentally for cold 
cesium
atoms when driven in the presence of gravity. A detailed theoretical 
explanation of 
this quantum effect is presented here. The theory makes use of invariance 
properties of
the system, that are similar to the ones of solids, leading to a separation 
into
independent kicked rotor problems.  The analytical solution makes use of a 
limit
that is very similar to the semiclassical limit, but with a small parameter 
that is
{\em not} Planck's constant, but rather the detuning  from the 
frequency  that is
resonant in absence of gravity.}

\end{abstract}

\vspace{1cm}


\section{Introduction}

The kicked rotor is a standard system used in the investigation of
 classical Hamiltonian chaos  and 
its manifestations in quantum mechanical systems
\cite{KR,chirikov,lichtenbergb}.
The motion is classically described by the 
Standard Map. The size of the chaotic 
component in the phase space of this map 
increases with the driving strength, 
and when the latter is sufficiently strong unbounded 
diffusion in action space takes place. Important deviations  
are however found for some values of the driving strength
\cite{chirikov,lichtenbergb,ZE}, due to the onset  of so-called 
accelerator modes, that produce linear, rather than diffusive, 
growth of momentum along orbits in a set of positive measure. 
Quantization imposes remarkable modifications. 
For typical parameter values, the classical diffusion is quantally 
suppressed  by a mechanism 
that is similar to Anderson localization 
in disordered solids \cite{KR,FGP}. 
The 
accelerator modes decay in time due to quantum 
tunneling \cite{hoa}. Quantum resonances are found when the natural 
frequency of the rotor is 
commensurate with the frequency of the driving \cite{IzShep}. 
The quasienergy states are then 
extended in angular momentum, leading to ballistic (i.e., linear) growth 
of the latter in time. 
  
The quantal
suppression of classical diffusive transport first 
observed  in the quantum kicked rotor is actually a more  
general phenomenon, now known as dynamical localization.
Theoretical predictions \cite{IEEE}
prompted the first experimental observations of this phenomenon
for microwave driven Hydrogen and Hydrogen 
like atoms\cite{expH}. 

The most direct experimental realization of the quantum kicked rotor is 
achieved 
in the field of atom optics, by a technique pioneered by Raizen and coworkers 
\cite{raizen1}.  
Laser-cooled Atoms (first Sodium and later Cesium)  
are driven  by application of a 
standing electromagnetic wave. The 
frequency of the wave  is slightly detuned from resonance, so a dipole 
moment is induced in  the atom. This moment 
couples with the driving field, giving rise 
to a net force on the center 
of mass of the atoms, proportional to the square of the electric field
\cite{GSZ,CT}. 
As the wave  is periodic in space, the atom is thus subjected to a periodic 
potential. The wave is turned on and off periodically in time, 
and the time it is on is 
much shorter then the time it is off. A realization of a periodically kicked 
particle is then obtained. In 
real experiments the duration of the kicks is always 
finite, setting a bound on the 
the momentum range 
wherein the $\delta$-kicked model is applicable. At very large 
momentum  the driving becomes 
adiabatic, leading to trivial classical and quantum 
localization in momentum
\cite{BFS,raizen1}. 

The basic 
difference between 
the kicked particle and the kicked rotor is that the momentum of the 
particle 
is  not discrete as is the angular momentum of the
rotor. As we shall review in Sect.(2.2) , this difference is circumvented 
by Bloch theory. The spatial periodicity of the driving 
only allows for transitions between 
momenta that differ by integer multiples of $\hbar G$, with $2\pi G^{-1}$ 
the spatial period of the driving potential. This  
implies conservation of quasi-momentum.  The 
particle wavepacket is a continuous superposition of 
states with given quasi-momenta. The dynamics at 
any one fixed quasi-momentum is 
that of a rotor, which however differs from the standard kicked rotor because 
of a constant shift in the angular momentum eigenvalues, proportional to the 
given quasi-momentum. This modification of the kicked rotor dynamics is 
formally the same as that produced by a 
Aharonov-Bohm flux threading the rotor.
It does not crucially affect dynamical localization \cite{phirot}, 
so the latter  carries over to the particle dynamics.
In experiments \cite{raizen1} it is found that for typical 
values of the parameters, an initial narrow gaussian distribution 
around zero momentum spreads into 
an exponential distribution, characteristic of 
dynamical or Anderson localization. 

The difference between the rotor's  and the particle's dynamics due to the 
presence of a continuum of quasi-momenta is indeed crucial  
in what concerns quantum resonances at ``commensurate'' values of 
the kicking period, because these only occur 
at special  values of quasi-momentum. Therefore, nearly all 
quasi-momenta  involved in a particle's  wave packet 
would not be in resonance \cite{raizen2}. In this paper we present an exact 
calculation, 
showing that the quadratic spread in momentum grows in this case 
linearly in time,  in contrast to the situation found for the kicked 
rotor, where this spread is quadratic in time.  A more detailed 
analysis 
of the kicked particle dynamics at resonance will be presented elsewhere
\cite{WGF}.

In the above discussed experiments, gravity had but negligible effects,   
as driving of the atoms took place in the 
horizontal direction. In 
recent experiments \cite{Ox1,Ox2,Ox3}, that provide the 
subject of the theoretical analysis of 
the present 
paper, atoms were driven in the vertical direction, 
and gravity was  
found to produce remarkable effects.  
In the vicinity of the resonant frequencies  
of the kicked rotor, a new type of ballistic spread in momentum 
was experimentally observed  \cite{Ox1,Ox2,Ox3}.  
A fraction of the atoms are steadily accelerated, at a rate which  
is faster or slower than the gravitational acceleration 
depending on what side of the 
resonance the driving frequency is. Such atoms 
are exempt from the diffusive spread that takes place for the other atoms, 
and  
their acceleration depends on the difference between the driving 
and the resonant frequencies. 
\footnote{The main experimental results are clearly presented in 
Figs 4 and 13 of Ref. \cite{Ox2}.} 
In ref. \cite{Ox2} a physical  explanation  
was given, and it was stressed  that  
the phenomenon is resemblant  of the accelerator modes 
in the Standard Map. The accelerating parts of the distributions were 
hence termed ``quantum accelerator modes'', at once emphasizing 
that resemblance, and their purely quantal nature; in fact,  they have 
no classical counterpart   in the 
classical dynamics of the kicked particles in gravity.\footnote{
For some  parameter values, the latter dynamics does exhibit accelerator 
modes, which are however unrelated to the experimentally observed ones.}

In this paper we present a theory that 
explains the experimental results in terms of the 
exact quantum equations of 
motion, 
and allows for  further predictions. 
The main result 
is that the quantum accelerator modes in presence of gravity do 
indeed correspond to accelerator modes  of a classical map. This map is 
not, however, the one given by the proper classical limit
$(\hbar\to 0)$. It emerges of a quasi-classical asymptotics, 
where the small parameter is not $\hbar$ but rather  
the detuning of the kicking frequency from the resonance of the kicked 
rotor. Though  a formally simple variant of the Standard Map, it is 
endowed with a rich supply of accelerator modes and complex 
bifurcation patterns.   

Our analysis starts from the time-dependent Schrodinger equation 
for the kicked particle in the laboratory frame. A simple gauge 
transform reduces the corresponding Floquet operator to that 
of a particle kicked by a potential, that  is quasi-periodic 
(and not just periodic) in space (Sect.(2.1)). The corresponding incommensuration 
parameter adds to the kicking period, in building a formidable 
mathematical problem.   
In particular, quasi-momentum is not 
conserved, preventing implementation of the Bloch theory. 
Moving to a free-falling frame removes this difficulty and restores 
decomposition into independent rotor problems,  at 
the price of time-dependent Floquet operators (Sect.(2.3)). On such operators 
we work out the mentioned quasi-classical approximation at small 
detuning from resonance.

Variants 
of the kicked-rotor, in which some parameter was allowed to change with 
time, 
have been considered earlier \cite{qpp}, 
the issue being what degree 
of uncorrelatedness in the time-dependence is sufficient to destroy 
localization. For the quasi-periodic dependence of the present model, 
this issue is as yet  unsolved, and  depends on 
incommensuration properties. In a mathematical aside of this work, 
we prove that in the presence of gravity and at resonant values of the 
kicking period the particle energy (in the falling frame) grows 
linearly in time, in sufficiently incommensurate cases at least.

\section{Discrete-Time Quantum Dynamics}

The dynamics of the atoms that are falling as a result of gravity and are 
kicked by the
external field is modeled by the time-dependent Hamiltonian:
\begin{equation}
\label{ham1}
{\hat H}(\ti)
=\frac{{\hat P}^2}{2M} -Mg{\hat X}+
\kappa \cos(G{\hat X})\sum\limits_{t=-\infty}^{+\infty}
\delta(\ti-t T),
\end{equation}
where $\ti$ is the continuous time variable, the integer 
variable $t$ counts the kicks, ${\hat P},{\hat X}$ are the momentum 
and the position operator respectively, $M$ is the mass of the atom, 
$2\pi G^{-1}$ is the spatial period of the kicks, $\kappa$ is the kick 
strength, and $T$ is the kicking period in time. The positive $x-$direction is 
that of the gravitational acceleration. 
Without changing notations, we rescale  
momentum ${\hat P}$ in units of $\hbar G$, position 
${\hat X}$ in units of $G^{-1}$, and mass in units of $M$. Then the    
energy $E$ comes in units of $\hbar^2G^2/M$, and 
time $\ti$ in units of $M/(\hbar G^2)$. The reduced Planck 
constant is equal to $1$, and the Hamiltonian takes  the following
 form:
\begin{equation}
\label{ham3}
{\hat H}(\ti)\;=\;\frac{{\hat P}^2}{2}-\frac{\eta}{\tau}{\hat X}
+k\cos({\hat X})\sum\limits_{t=-\infty}^{+\infty}\delta(\ti-t\tau)\;,
\end{equation}
where:
\begin{equation}
\label{par}
k\;=\frac{\kappa}{\hbar}\;\;\,\;\;\tau\;=\;\frac{\hbar TG^2}{M}
\;\;,\;\;\eta\;=\;\frac
{MgT}{\hbar G}\;,
\end{equation}
In the above defined units, $\eta/\tau$ is the gravitational acceleration.
The dynamics 
is  fully characterized by the dimensionless parameters $k,\tau,\eta$.

In the following Dirac notations will be used: e.g., 
$\psi(x)=\langle x|\psi\rangle$ and $\psi(p)=\langle p|\psi\rangle$ will 
denote the wave function in the position and in the momentum representation respectively.

\subsection{Floquet operators.}

The  quantum evolution over  a sequence 
of discrete times spaced by one period of the external periodic driving 
is obtained by repeated 
application of the Floquet operator. This is the single unitary operator 
which gives the evolution from  any instant to the next one in the given 
sequence of times. Our discrete times will be the kicking times themselves 
\footnote{Other choices lead to different Floquet operators, which are 
nonetheless unitarily equivalent to the present one.}. 
{\it Throughout the following, time is a discrete variable, given by 
the kick counter $t$.}
As the state discontinuously changes at the kicking times, we further 
specify the 
state  at time $t$ to be the one immediately {\it after  } the $t-$th 
kick. 
The Floquet 
operator $\Uop$ is then found by integrating the Schr\"odinger equation (with 
the 
Hamiltonian (\ref{ham3})) from t$=0_+$ to t$=\tau_+$: 
\begin{equation}
\label{ini}
\Uop\;=\;\Kop{\hat F}\;=\;e^{-ik\cos({\hat X})}{\hat F}\;,
\end{equation}
where $\Kop$ describes the kick, and ${\hat F}$ describes free fall 
inbetween kicks. 
 With the present  units, 
the energy eigenfunctions $u_E(p)=\langle p|E\rangle$ of the 
particle  in the gravity field read \cite{LL} :
$$
u_E(p)\;=\;\left(\frac{\tau}{2\pi\eta}\right)^{1/2}\;e^{i\frac{\tau}{\eta}
(Ep-\frac{p^3}{6})}\;.
$$
Therefore, apart from a constant phase factor,
\begin{eqnarray}
\langle p'|{\hat F}|p''\rangle\;&=&\;\int dE\; e^{-iE\tau }u_E(p')u^*_E(p'')
\nonumber\\
&=&\;\delta(p'-p''-\eta)\;e^{-i\frac{\tau}{2}(p'-\frac{\eta}{2})^2}\nonumber
\end{eqnarray}
Moreover,
$$
\langle p|e^{-ik\cos({\hat X})}|p'\rangle\;=\;\sum\limits_{n=-\infty}
^{\infty}K_n\;\delta(p-p'-n)\;
$$
where $K_n=(-i)^nJ_n(k)$ and $J_n(k)$ are the Bessel functions of the 1-st kind.
Replacing in (\ref{ini}) we get the explicit form of the propagator in the 
laboratory frame:
\begin{equation}
\label{lab}
(\Uop\psi)(p)\;=\; \sum\limits_{n=-\infty}
^{\infty}K_n\;e^{-i\frac{\tau}{2}(p-n-\frac{\eta}{2})^2}\
\psi(p-n-\eta)\;.
\end{equation}
Note that $K_n=K_{-n}$. 
A more transparent formulation is gained by introducing  the operators:
$$
\Rop=e^{-i\tau {\hat P}^2/2}\;\;,\;\;\Sop=e^{i\eta {\hat X}/2}
$$
One may then write: 
\begin{equation}
\label{op}
\Uop=\Sop\Kop\Rop\Sop
=\Sop^{\dagger} \Uop'\Sop\;\;\;,\;\;\;
\Uop'=\Sop^2 \Kop\Rop
\end{equation}
Thus $\Uop$  only differs by a unitary transformation  (in fact a gauge 
transformation) from $\Uop'$, which 
has the simple form: 
\begin{equation}
\label{fin}
\Uop'\;=\;= e^{i(\eta {\hat X}-k\cos({\hat X} ))}\;e^{-i\tau {\hat P}^2/2}
\end{equation}
Thus formulated, the problem is that of a particle freely moving  on a line, 
except for time-periodic kicks. The spatial dependence of the kicks 
is periodic when $\eta$ is rational, quasi-periodic otherwise.

\subsection{Quasi-momentum and  Kicked Rotors.}

If  $\eta=0$, 
then (\ref{fin}) is   formally similar to the Kicked Rotor model 
\cite{KR}, from which it however differs in one crucial respect:
whereas the latter has the kicked particle moving on a circle, 
(\ref{fin}) has the particle moving on a line instead. 
A link  between the two models is established by 
the spatial periodicity of the kicking potential.  
We review this well-known 
construction, because it plays a fundamental role in this paper.

At $\eta=0$ the evolution operator $\Uop$ 
commutes with spatial translations by multiples of $2\pi$. As is well 
known from Bloch theory,
this enforces 
conservation of {\it quasi-momentum}. In our units, this is 
given by the fractional part of 
momentum, and will be denoted by  $\beta$. We then introduce a family 
of fictitious rotors (particles moving on a circle) parametrized by 
$\beta\in [0,1)$, with angle 
coordinate $\theta$ (henceforth named $\beta-${\it rotors}), 
and denote $|\Psi_{\beta}\rangle$ their states. 
For integer $n$ we denote $|n\rangle$ 
the angular momentum eigenstates of these rotors so that in the $\theta-$
representation $\langle\theta|n\rangle=(2\pi)^{-1/2}\exp(in\theta)$. 
To states $|\psi\rangle$ of the particle we associate states 
$|\Psi_{\beta}\rangle$ of the $\beta-$rotors as follows:
\begin{equation}
\label{psibeta}
\langle\theta|\Psi_{\beta}\rangle\;=\;\frac{1}{\sqrt{2\pi}}
\sum\limits_{n}\langle n+\beta|\psi\rangle e^{in\theta}\;.
\end{equation}
In the angular momentum representation,
\begin{equation}
\label{psin}
\langle n\vert\Psi_{\beta}\rangle\;=\;\langle n+\beta|\psi\rangle
\end{equation}
Note that $|\Psi_{\beta}\rangle$ is not necessarily normalized to $1$, 
even if $|\psi\rangle$ is. 
Conversely, the state of the particle is retrieved from the $\beta-$rotor
states via
\begin{equation}
\label{rec}
\langle p|\psi\rangle\;=\;\frac{1}{\sqrt 
{2\pi}}\int_0^{2\pi}d\theta\;\langle\theta|
\Psi_{\beta}\rangle\;e^{-in\theta}\;\;,\;\;n=[p]\;\;,\beta=\{p\}\;.
\end{equation}
where $[p],\{p\}$ denote the integer and the fractional part of the 
momentum $p$ respectively. Using (\ref{rec}) and Poisson's summation 
formula, one obtains the wave function in the $x-$representation: 
\begin{equation}
\label{recx}
\langle x|\psi\rangle =\frac{1}{\sqrt{2\pi}}
\int dp\;e^{ipx}\langle p|\psi\rangle=\int_0^1d\beta\;e^{i\beta x}
\langle x\;\mbox{\rm mod}(2\pi))|\Psi_{\beta}\rangle\;.
\end{equation}
Fixing a sharp 
value of $\beta$ yields a spatially  extended   
 state (a Bloch wave ) for the particle, even with a normalizable 
rotor wave function.

It follows from (\ref{lab}) and (\ref{psin}) (with $\eta=0$) that as $|\psi\rangle$ evolves 
into 
$\Uop^{t}|\psi\rangle\;$, 
$|\Psi_{\beta}\rangle$ in turn evolves into $\Up_{\beta}^{t}|\Psi_{\beta}
\rangle$, 
where:
\begin{equation}
\label{rotdyn}
\Up_{\beta}=(\Kp\Rp_{\beta})\;\;,\;\;
\langle n|\Kp|m\rangle=K_{n-m}\;\;,\;\;\langle n|\Rp_{\beta}
|m\rangle
=\delta_{nm}\;e^{-i\frac{\tau}{2}(\beta+m)^2}\;.
\end{equation}
Eqn.  (\ref{rotdyn})
may also be written as follows:
\begin{equation}
\label{Ulab}
\Up_{\beta}\;=\;e^{-ik\cos({\hat \theta})}\;e^{-i\frac{\tau}{2}(\Np+\beta)^2}.
\end{equation}
where $\Np=\sum_n n|n\rangle\langle n|$ is the angular momentum operator:
$\Np=-i\frac{d}{d\theta}$ in the $\theta-$representation,
At $\beta=0$, eqn. (\ref{Ulab}) is the standard Floquet operator of the 
Kicked Rotor. At $\beta\neq 0$ one obtains a variant of the Kicked 
Rotor, which has also been studied, $\beta$ being typically given 
the physical meaning of an external magnetic flux\cite{phirot}. 

 We have thus shown that in the presence of conserved quasi-momentum the 
particle  dynamics can be determined as follows: 
given an initial particle (pure) state $|\psi(0)\rangle$ one 
 first computes the corresponding rotor states $|\Psi_{\beta}(0)\rangle $
 as above described. These separately evolve 
into $|\Psi_{\beta}(t)\rangle$ at time $t$. The particle state $
|\psi(t)\rangle$ is  
finally reconstructed using (\ref{rec}).  
The process of decomposing the particle dynamics 
in a bundle  of rotors  will 
henceforth be named ``the Bloch-Wannier fibration''. It is the 
quantal counterpart of the classical process of ``folding back'' 
the particle trajectory onto a circle, by taking the 
$x-$coordinate mod$(2\pi)$.
 
The kinetic energy of the particle  
 at time $t$ is:
\begin{eqnarray}
\label{mess}
\En(t)\;&=&\;\frac{1}{2}\int dp\;p^2|\psi (p,t)|^2\;=\;
\frac{1}{2}\sum\limits_{n=-\infty }^\infty
\;\int_0^1d\beta\;(n+\beta)^2|\psi (n+\beta ,t)|^2\nonumber\\ 
&=&
\frac{1}{2}\int_0^1d\beta\int_0^{2\pi }d\theta\left \{ \left| \frac d{d\theta
}\langle \theta|\Psi_{\beta}(t)\rangle\right| ^2
+\beta ^2|\langle\theta|\Psi _{\beta}(t)\rangle|^2+\right. \nonumber\\
&-&2i\left. \beta\langle
\Psi_{\beta}(t)|\theta\rangle 
\frac d{d\theta} \langle\theta|\Psi_{\beta} (t)\rangle\right \}
\end{eqnarray}
In the case of unbounded propagation in momentum space, 
it is the 1st term
within the curly brackets which yields the dominant contribution at large $t$
(the 2-nd is
bounded by $1$, the 3-d is order of square root of the 1-st.). Note that 
the 1-st  term is 
the average over $\beta$ of the $\beta-$ 
rotors kinetic energy.

\subsection{Motion in the falling frame.}

In the case of nonzero gravity, 
quasi momentum is not 
conserved any more because the kicking potential in (\ref{fin}) is not 
periodic 
(unless $\eta$ is rational). However, 
a quasi-momentum -conserving evolution is restored 
by the substitution :
$$
\psi_f(p,t)\;=\;\langle p+t\eta|\Uop^{t}|\psi),
$$
Since $\eta$ is the momentum gained over one period  due to gravity, 
this amounts to measuring momentum in the free-falling frame (recall 
that the positive $x-$direction is that of gravitational acceleration).

From (\ref{lab}) it follows that 
\begin{equation}
\label{fall}
\psi_f(p,t+1)\;=
\sum\limits_{n} K_{n}\;
e^{-i\frac{\tau}{2}(p-n 
+t\eta+\frac{\eta}{2})^2} \psi_f(p-n,t)\;.
\end{equation}
that may be rewritten as :
\begin{equation}
\label{fall7}
|\psi_f(t+1)\rangle=\Uop_f(t)|\psi_f(t)\rangle\;
\;,\;\;\Uop_f(t)=e^{-ik\cos({\hat X})}e^{-i\frac{\tau}{2}({\hat P}+t\eta+\frac{\eta}{2})^2}\;.
\end{equation}
The operator ${\hat U}_f(t)$  describes up to a constant phase the
evolution 
from (continuous) time $\ti=(t\tau)_{+}$ to time $(\ti+\tau)_{+}$ under 
the time-dependent Hamiltonian 
\begin{equation}
\label{ham2}
{\hat H}_f(\ti)\;=\;
\frac{1}{2} (\hat{P}+\frac{\eta}{\tau} \ti)^2+k\cos (\hat{X})\sum\limits_
{t=-\infty}^{+\infty}
\delta(\ti-t\tau)\;,
\end{equation}
that describes the motion in the falling frame. This Hamiltonian 
is related to that in eqn.(\ref{ham3})
by the gauge transformation $e^{i\eta{\hat X}\ti/\tau}$.
The classical dynamics corresponding to (\ref{fall7}) is given by the 
area-preserving, time-dependent map 
in the $(x,p)$ plane:
\begin{equation}
\label{class}
p_{t+1}=p_t+k\sin(x_{t+1})\;\;,\;\;x_{t+1}=
x_t+\tau(p_t+t\eta+\frac{\eta}{2})\;.
\end{equation}
Since $\hbar=1$ in our units, the classical limit is approached as  
\begin{equation}
\label{clim}
k\rightarrow\infty\;,\;\tau\rightarrow 0\;,\;\eta\rightarrow\infty\;,
\;k\tau\rightarrow\mbox{\rm const.}\neq 0\;,\;
\eta\tau\rightarrow\mbox{\rm const.}\neq 0\;.
\end{equation}
The evolution (\ref{fall}) only mixes momenta which differ  
by integers: hence,  quasi-momentum is 
conserved, so the Bloch-Wannier fibration can be fully implemented, as 
described in 
the previous Section. 
Eqn.(\ref{rotdyn}) now becomes: 
\begin{equation}
\label{rotdyng}
|\Psi_{\beta}(t)\rangle\;=\;
\prod\limits_{r=0}^{t-1}\;\Up_{\beta}(r)|\Psi_{\beta}\rangle\;,\;\;\;
\Up_{\beta}(r)=\Kp\Rp_{\beta}(r)
\end{equation}
(all operator products ordered from right to left),
where
\begin{equation}
\label{rotopg}
\langle n|\Kp|m\rangle=K_{n-m}\;\;,\;\;\langle n|\Rp_{\beta}(r)
|m\rangle
=\delta_{nm}\;e^{-i\frac{\tau}{2}(\beta+m+r\eta+\eta/2)^2}\;.
\end{equation}
Consequently, in the falling frame, 
\begin{equation}
\label{Ufall}
\Up_{\beta}(t)\;=\;e^{-ik\cos({\hat\theta})}\;e^{-i\frac{\tau}{2} (\Np+
\beta+\eta
t+\eta/2)^2}.
\end{equation}

\section{Features of Quantum Dynamics in the falling frame.}

The theoretical importance of the  Kicked Rotor model lies with its asymptotic 
properties in time. Best known among these are 
dynamical localization, and quantum resonances.
Though not crucial to current experiments, the  long-time asymptotics   
is an important theoretical question for the present models, too. 
Apart from resonances, we do not  
attempt at a thorough  theoretical 
analysis of this issue, which is 
likely  to critically depend on  the arythmetic type of $\tau,\eta,\beta$. 
Though some results in this respect will be presented later, 
we mostly focus on intermediate-time features of the quantum dynamics, 
directly connected to experimental findings.

In this section the falling frame is constantly used, without any further 
reference to the laboratory frame. The suffix 
$_f$ will be omitted in order to  simplify notations.

\subsection {Resonances}

\subsubsection {Zero Gravity}

If the kicking period $\tau$ is commensurate to $4\pi$, that is 
$\tau=4\pi p/q$ with $p,q$ relatively prime integers, and in addition 
$\beta=m/(2p)$ with $m<2p$ an integer,  
then the $\beta-$Kicked Rotor 
exhibits the phenomenon of Quantum Resonance \cite{IzShep}. 
In that case, indeed, the rotor dynamics commutes with translations 
{\it in momentum} by multiples of $q$. This typically  
results in band (absolutely continuous) quasi-energy spectrum 
and ballistic spread of the rotor wave packet in the momentum representation.
For special values of $q$ the bands may however be flat. This is the case 
when $q=2$; the ballistic spread is then only observed at $\beta=1/2$,  
while at $\beta=0$ the dynamics is sharply localized in 
momentum (``anti-resonance''), as can be seen from (\ref{Ulab}), making use of 
the fact that 
$e^{-i\pi ln^2}=e^{-i\pi ln}$ holds for integer $l$ and $n$. 
The width of the quasi-energy
bands rapidly decreases as the order $q$ of the resonance 
increases\cite{IzShep}. 
Ballistic motion is then observed only after quite long times. 
This 
places high-order resonances beyond experimental observation. Our 
discussion is mostly focused on the main resonances ($q=1,2$). 

Since  the  quadratic growth of energy at resonant values of $\tau$ 
requires a sharp value of quasi-momentum 
(hence an extended particle state in position), 
it cannot  be observed in the 
 particle {\it wave packet} long-time dynamics, not even  at resonant values 
of $\tau$. 
Nevertheless, with a generic choice of the initial wavepacket, 
a resonant effect   is still manifest, in the form of  
{\it linear} asymptotic growth of the particle energy. 
 This is qualitatively understood as follows.  
 At time $t$, rotors whose 
quasi-momenta lie within $\sim 1/t$ of the resonant value(s) are still 
mimicking the resonant growth of energy $\propto t^2$. 
Assuming a smooth initial 
distribution of quasi-momenta, such rotors enter the average 
over $\beta$ (\ref{mess}) with a weight  $\propto 1/t$.  
This directly 
leads to the linear growth of $\En(t)$. The latter can also be 
rigorously proven, and the proportionality factor computed. This is done 
in the Appendix, under the hypothesis that the initial wave function 
of the particle 
is such that $\langle\psi|{\hat X}^{2\alpha}|\psi\rangle<\infty$ for some 
$\alpha>1$.

\subsubsection{Nonzero gravity}

The $\beta-$rotor dynamics at resonant values of $\tau$ in the 
presence of gravity is a nontrivial mathematical problem, which may give 
rise to different types of quantum transport depending on the arythmetic 
type of $\eta$.

In the Appendix we prove that for $\tau=2l\pi$ ,($l$ integer),  and 
for all irrational $\eta$ in a set of full Lebesgue measure,
the energy $ \En(t)$ (in the falling frame) 
grows  like $k^2t/4$, under the hypothesis that the initial wave 
function satisfies $\langle\psi|{\hat X}^{2\alpha}|\psi\rangle<\infty$
for some $\alpha>1$.

\subsection {Accelerator Modes}

\subsubsection{Quantum Rotor Dynamics  near Resonance.}

We shall now analyze the quantum dynamics at values of $\tau$ close to the 
resonant values $2\pi l$ ($l>0$ integer) and for large kicking strength $k$. 
We hence assume $\tau=2\pi l+\epsilon$ with $l$ integer and $|\epsilon|<<1$, 
and  rescale  $k={\tilde k}/|\epsilon|$. 
Replacing in (\ref{rotopg}) and 
noting $e^{-i\pi ln^2}=e^{-i\pi ln}$ we get:
$$
\langle m|\Rp_{\beta}(t)|n\rangle\;=\;
\delta_{mn}\;e^{-i\frac{\tau}{2}(\beta+t\eta+\eta/2)^2}
\;e^{-i(\epsilon\frac{ n^2}{2}
+n(\pi l+\tau\beta+\tau t\eta+\tau\frac{\eta}{2}))}
$$
whence it follows that (apart from an irrelevant  phase factor)
\begin{equation}
\label{epsdin1}
\Up_{\beta}(t)\;=\;e^{-\frac{i}{|\epsilon|}{\tilde k}\cos{\hat \theta}}\;
e^{-\frac{i}{|\epsilon|}\Hp_{\beta}({\hat I},t)}
\end{equation}
where\footnote{The integer time variable $t$ is fixed on both sides, and the 
2nd exponential on the right hand side is that of a constant operator. } 
\begin{equation}
\label{epsdin2}
{\hat I}=|\epsilon|\Np=-i|\epsilon|\frac{d}{d\theta}\;\;\;,\;\;
\Hp_{\beta}({\hat I},t)=\frac12\; \mbox{\rm sign}(\epsilon) {\hat I}^2+
{\hat I}(\pi l+\tau (\beta+t\eta+\frac{\eta}{2}))
\;,
\end{equation}
If $|\epsilon|$ is assigned the role of Planck's constant, then 
 (\ref{epsdin1}),(\ref{epsdin2}) is the formal quantization 
of either of the  following classical (time-dependent) maps :
\begin{equation}
\label{clmap}
I_{t+1}=I_{t}+{\tilde k}\sin (\theta_{t+1})\;\;,\;\;
\theta_{t+1}=\theta_t\pm I_t+\pi l+\tau(\beta+t\eta+\eta/2)\;\mbox{\rm mod}
(2\pi)\,
\end{equation}
where $\pm$ has to be chosen according to the sign of $\epsilon$.\footnote{
The double sign would not appear if $\epsilon$ were used 
as the ``Planck constant'',  
rather than $|\epsilon|$. In this way the sign of $I$ would be 
reversed with respect to that of physical momentum whenever $\epsilon<0$.}
The small $|\epsilon|$ asymptotics of the quantum 
$\beta-$ rotor is thus equivalent to a  
quasi-classical approximation   based on the ``classical'' 
dynamics  (\ref{clmap}). 
We emphasize that ``classical'' here is not  
related to the $\hbar\to 0$ limit but to the limit $\epsilon\to 0$ instead.
The two limits are actually incompatible with each other except possibly 
when $l=0$ (see 
(\ref{clim})). 
For the sake of clarity the term ``$\epsilon-$classical'' will be used 
in the following.

\subsubsection{$\epsilon-$Classical rotor dynamics.}

\begin{figure}
\centerline{\epsfxsize=14cm\epsfysize=8cm\epsffile{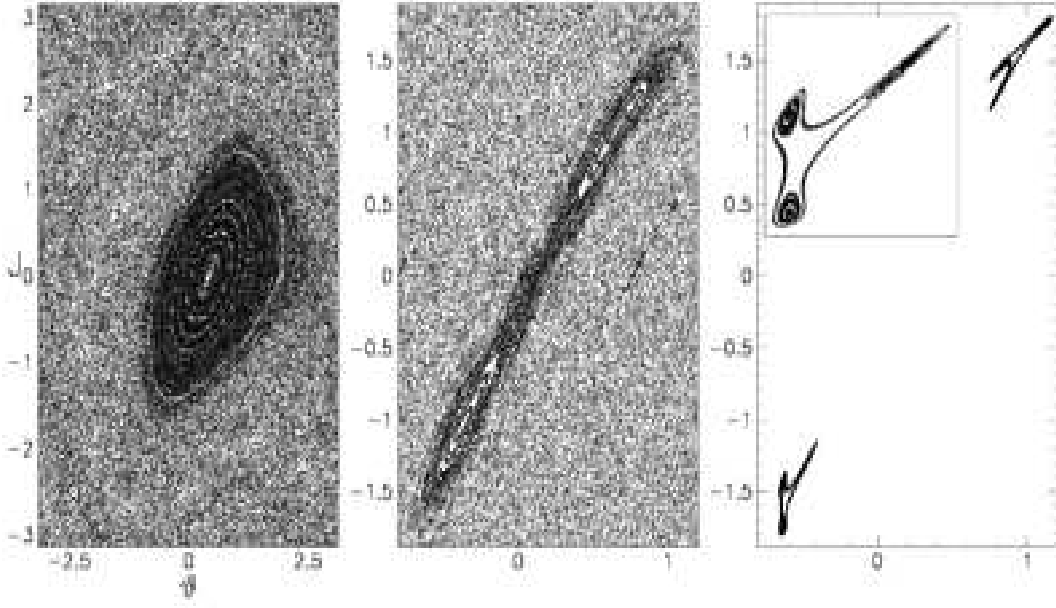}}
\caption{\small Phase portraits for the map (\ref{clmap1}) on the 2-torus,  
with 
$\tau=5.86,\eta/\tau=0.01579,\epsilon=-0.423$. Here  
the torus is mapped onto $[-\pi,\pi)\times[-\pi,\pi)$. 
One stable fixed point with $\jG=0$ exists for $0.542<\kcl<
4.037$.
Left: $\kcl=1.329$, the stable fixed point is at $J^*=0$, $\theta^*=0.42$.
Center: $\kcl=4.232$,
a stable period-2, $\jG=0$ orbit is visible. Right: $\kcl=4.494$, the 
 period-$2$ orbit  has left room to 
a stable  period-$6$ orbit. The lower left part is magnified in the inset.
 }
\label{fig1}
\end{figure}

Changing variable 
to $J_t=I_t\pm \pi l\pm\tau (\beta+\eta t +\eta/2)$ removes the explicit 
time dependence of the maps (\ref{clmap}), yielding:
\begin{equation}
\label{clmap1}
J_{t+1}=J_t+{\tilde k}\sin (\theta_{t+1})\pm\tau\eta\;,\;\;
\;\theta_{t+1}=\theta_t\pm J_t\;.
\end{equation}
These area-preserving maps are $2\pi$- periodic in  $J$ and $\theta$, 
so they define smooth, area-preserving 
maps  of the 2-torus parametrized by 
$\jt=J$ mod$(2\pi)$, $\vartheta=\theta$ mod$(2\pi)$. Such ``toral maps'' 
are conjugated to each other under $\jt\to 2\pi-\jt$ mod$(2\pi)$, 
$\vartheta\to\vartheta +\pi$ mod$(2\pi)$. They differ from the Standard 
Map by the constant drift $\eta\tau$. 

Let $\jt_0,\vartheta_0$ be a period-$\pG$ fixed point of either  toral map. 
Iterating (\ref{clmap1}) with $J_0=\jt_0\;,\;\theta_0=\vartheta_0$, at $ 
t=\pG$ one obtains:
\begin{equation}
\label{fixp} 
J_{\pG}=J_0+2\pi \jG\;\;,\;\;\theta_{\pG}=\theta_0+2\pi n
\end{equation} 
for some integers $\jG\;,\;n$. The points $(\jt_t,\vartheta_t)$ 
with $0\leq t\leq \pG-1$ are period-$\pG$ fixed points  
themselves, and define a period-$\pG$ periodic orbit of the toral 
map, which is primitive if all such points are distinct.

Period-$1$ fixed points are given on the 2-torus 
by $\jt_0=0$, $\vartheta_0=\theta_{\jG}$, where
\begin{equation}
\label{fix}
\sin(\theta_{\jG})\;=\;\frac{2\pi\jG\mp\tau\eta}{\kcl}
\end{equation}
and $\jG$ is any integer such that:
\begin{equation}
\label{ex}
-\kcl\pm\tau\eta\leq 2\pi\jG\leq \kcl\pm\tau\eta\;.
\end{equation}
No such point exists if $\kcl<\tau\eta<\pi$; 
two (at least) exist  (that is, (\ref{fix}) is solvable for at least one value 
of the integer $\jG$) whenever 
${\tilde k}\geq\pi$.  Two period-$1$ fixed points 
with $\jG=0$ exist whenever $\kcl\geq\tau\eta$. 
In order for the  period-$1$ fixed  points (\ref{fix}) to be stable it is required 
that:
\begin{equation}
\label{stab}
-4\;<\;\pm {\tilde k}\cos (\vartheta_0)\;<\;0
\end{equation}

\begin{figure}
\centerline{\epsfxsize=14cm\epsfysize=8cm\epsffile{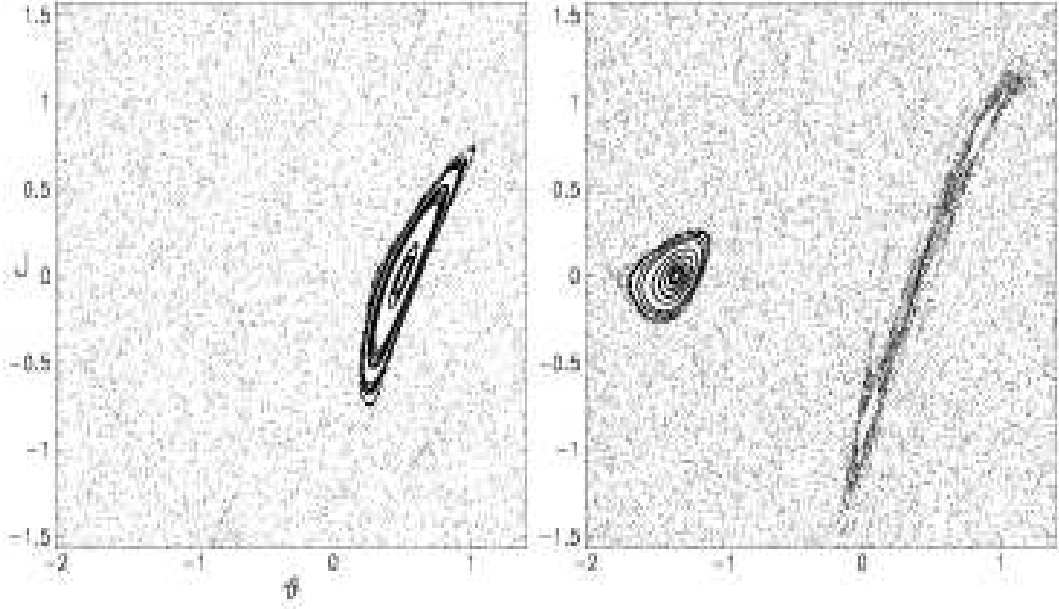}}
\caption{\small Same as Fig.\ref{fig1}, for $\eta/\tau=0.01579$, $\kcl=
0.8\pi(4\pi-\tau)$, and different values of $\tau$. Left: 
at $\tau=10.996$, one stable fixed point exists with period $1$, $\jG=0$.
On decreasing  $\tau$ it turns unstable at $\tau\approx 10.813$, 
generating a period-2 stable orbit. 
At $\tau\approx10.801$, one stable orbit  with period $1$, $\jG=-1$ appears. 
Both orbits are shown in the phase portrait on the right, drawn 
at $\tau=10.744$.}

\label{fig2}
\end{figure}

 From  (\ref{ex}), (\ref{stab}) it follows that for any 
integer $\jG$ each map (\ref{clmap1}) has exactly one 
stable period-$1$ fixed point on the 2-torus,  
given by (\ref{fix}) if, and only if,
\begin{equation}
\label{stabcond}
{\kcl}^{(\jG,\pm)}_{min}\;<\;\kcl\;<\;
{\kcl}^{(\jG,\pm)}_{max}\;\;,\;\;
{\kcl}^{(\jG,\pm)}_{min}=|2\pi\jG\mp\tau\eta|\;,\;
{\kcl}^{(\jG,\pm)}_{max}=\sqrt{16+(2\pi\jG\mp\tau\eta)^2}\;.
\end{equation}
At  $\kcl={\kcl}^{\jG,\pm}_{max}$ such  fixed points turn unstable and   
bifurcations occur. This is shown in Fig.\ref{fig1} for a case with $\jG=0$. 
At $\kcl={\kcl}^{0,-}_{max}=4.037$  a stable  period-$2$ orbit 
appears. 
This  becomes in turn unstable at $\kcl\approx 4.490$, 
and  a stable period-$6$ 
orbit is left . 
The size of the islands rapidly decreases through the 
sequence of bifurcations. At $\kcl=5$ no significant stability 
islands are any more  visible; at $\kcl=5.741$  and 
at $\kcl=6.825$ stable period-1 points with $\jG=-1$ and $\jG=1$ 
respectively appear. The rise and fall of these, and  of 
subsequent higher-$\jG$ period-1 points as well, are ruled by  
(\ref{stabcond}). 
Further examples of period-1 fixed points, and bifurcations thereof, 
are illustrated in Fig.\ref{fig2}.

\begin{figure}
\centerline{\epsfxsize=14cm\epsfysize=8cm\epsffile{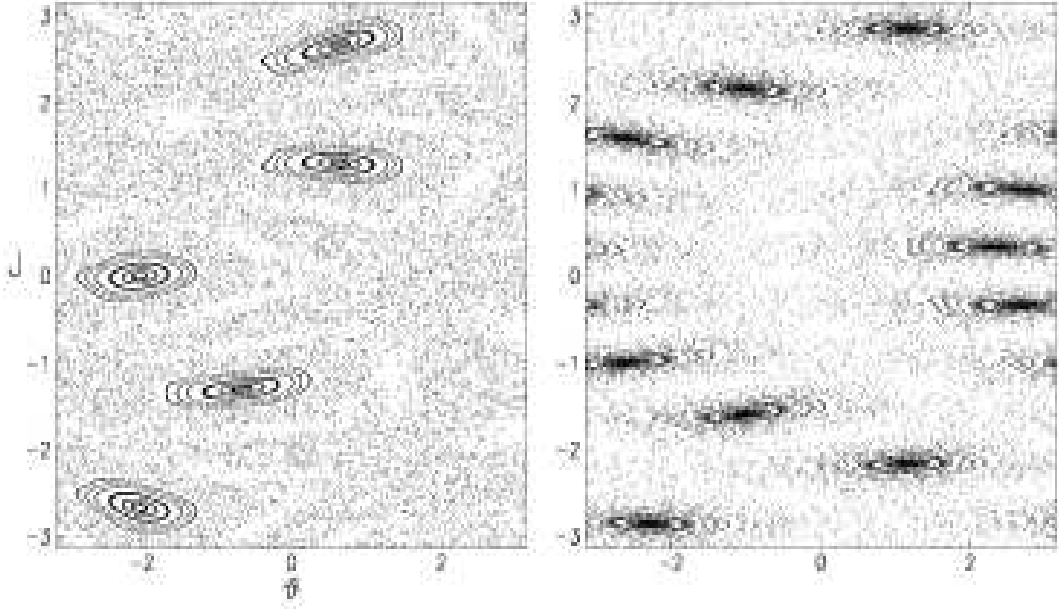}}
\caption{\small Phase portraits for the $\epsilon$-classical dynamics 
on the torus, showing a $(5,-2)$ periodic orbit at $\tau=12.472$, $\kcl=
0.8\pi(4\pi-\tau)$, 
$\eta=0.01579\tau$ (left) and a $(10,1)$ periodic orbit at  
$\tau=6.31$, $\kcl=0.8\pi(\tau-2\pi)$, 
$\eta=0.01579\tau$ (right).
}
\label{fig3}
\end{figure}

Examples of primary (that is, 
not born of period-1 points by bifurcation)
higher-period stable orbits 
are shown in Fig.\ref{fig3}. 
Generally speaking,  
the presence of {\it two} independent parameters ($\kcl$ and $\tau\eta$)
in the maps  
produces a remarkable variety of stable periodic orbits of higher periods, 
depending on the parameter values in complicated ways. 
Figs.\ref{fig1}, \ref{fig2} and \ref{fig3}
provide but a partial view of such complexity. They were singled out 
because they pertain to experimentally relevant parameter ranges; see the 
discussion in sect.(3.2.6). In particular, the value 
$\eta/\tau=0.01579$ constantly used in this paper is that 
of experiments 
in \cite{Ox1,Ox2,Ox3}.


\subsubsection{$\epsilon-$Classical Accelerator Modes.}

Primitive periodic orbits  of the toral maps yield families of  
{\it accelerator orbits} 
of the original dynamics (\ref{clmap}), marked by linear average growth 
of momentum with time:
\begin{equation}
\label{growth}
\theta_{\pG t}=\theta_0=\vartheta_0\;\mbox{\rm mod}(2\pi)\;\;,\;\;
I_{t\pG}=I_0+a\pG t\;, 
\end{equation}
where 
\begin{equation}
\label{accmod1}
I_0=\jt_0\mp\pi l\mp\tau(\beta+\frac{\eta}{2})+2\pi m\;\;,\;\;
a=\frac{2\pi\jG}{\pG}\mp \tau\eta\;.
\end{equation}
with $m$ any integer, and $\jt_0,\vartheta_0$ a period-$\pG$ fixed point. 
If such  accelerator orbits
are stable, then they are surrounded by 
islands  of positive measure in phase space, also leading to ballistic 
(linear) average growth  of momentum in time. These are named 
{\it accelerator modes}. 

We shall classify accelerator modes according 
to their {\it order} $\pG$ and {\it jumping index} $\jG$. By 
a $(\pG,\jG)$-accelerator mode we shall mean a mode, whose order and jumping 
index are given by the integers $\pG,\jG$ respectively.

\subsubsection{Quantum Accelerator Modes in  Rotor Dynamics.}

Initial physical momenta $n_0=|\epsilon|^{-1}I_0$ for $\epsilon-$classical 
accelerator modes are obtained from (\ref{accmod1}) 
 for any $0\leq \beta<1$ :
\begin{equation}
\label{accmod2}
n_0\;=\;\frac{2\pi m+\jt_0}{|\epsilon|}-\frac{\pi l}{\epsilon}
-\frac{\tau}{\epsilon}(\beta+\frac{\eta}{2})
\end{equation}
If the stable 
islands associated with $\epsilon-$classical accelerator modes have a large 
area compared to $|\epsilon|$, then they 
support  a large number  of quantum states. Thus they  
may trap some of the rotor's wave packet and give rise to 
{\it quantum accelerator modes}  traveling  in physical momentum 
space with speed $\sim a/|\epsilon|=-\tau\eta/\epsilon+2\pi\jG/(\pG |\epsilon|)$.
In order that such modes
may be observed, the phase space distribution associated with the initial 
rotor state must significantly overlap the islands. Even in that case the 
modes will eventually decay due to quantum tunneling out of the classical 
islands. 

\begin{figure}
\centerline{\epsfxsize=14cm\epsfysize=8cm\epsffile{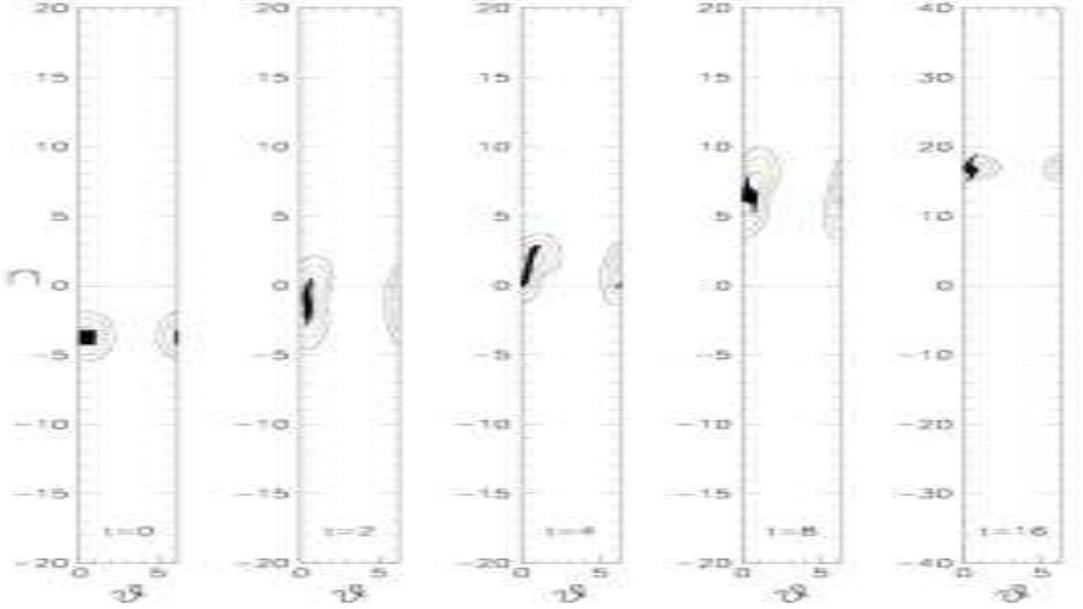}}
\caption{\small Quantum phase space evolution of the $\beta-$rotor with 
$\beta=0.218769$, $\kcl=1.329$, $\tau=5.86$, $\eta=0.093$ 
initially prepared in the  coherent state centered 
at the $\epsilon-$classical $(1,0)$ accelerator mode 
$n_0=-3.754$, $\theta _0 =0.420$. 
Contour plots of the Husimi functions of the rotor are shown at  
times $t=$$0$,$2$,$4$,$8$,$16$. 
The black spots in the centers of the contours are an ensemble of 
classical phase points, initially distributed in a square 
of area $\hbar=1$ centered at the mode. They evolve according to 
the $\epsilon-$classical dynamics (\ref{clmap}).}
\label{fig4}
\end{figure}

This picture is confirmed by numerical simulations. Fig.\ref{fig4}
 shows the quantum 
phase-space evolution  of a $\beta-$rotor 
started in a coherent state centered at the position of the 
$(1,0)$- accelerator mode generated by the fixed point shown in Fig.\ref{fig1}
 (left). 
 The Husimi functions  
computed at subsequent times  closely  follow the motion of the 
$\epsilon-$classical mode. 

\begin{figure}
\centerline{\epsfxsize=8cm\epsfysize=8cm\epsffile{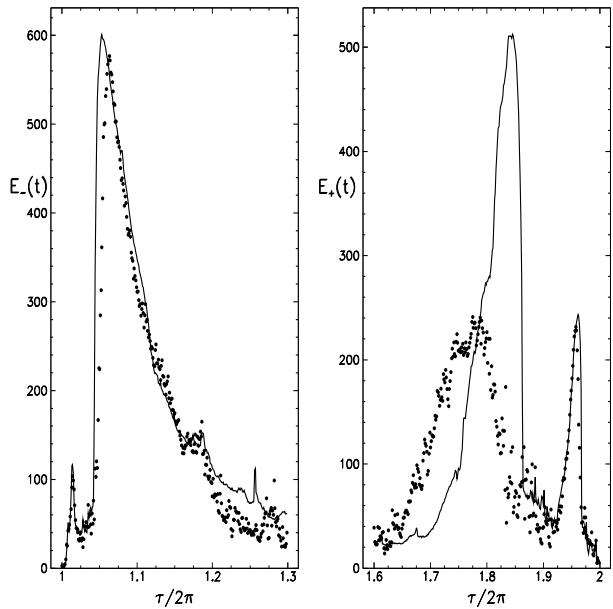}}
\caption{\small Full dots: Quadratic spread  
$E_{\pm}(t)$ of the quantum $\beta$-rotor over positive (right) 
and negative (left) momenta
after 
$30$ kicks, vs $\tau$, with  
$\eta/\tau=0.01579$, $k=0.8\pi$, $\beta=l\pi/\tau-\eta/2$,
$l=1$ (left) and $l=2$ (right).
The rotor was started at $n=0$. Full lines: same for 
an ensemble of $5\times 10^6$ 
classical rotors evolving according to (\ref{clmap}), started with  $n=0$ 
and $\theta$ uniformly distributed in $[0,2\pi]$. 
The exact meaning of $E_{\pm}(t)$ is explained in the text. At values 
of $\tau/2\pi \simeq 1.72$ 
the classical 
phase portraits are as shown in Fig.2.  } 
\label{fig6}
\end{figure}

In Fig.\ref{fig6} we compare quantum and $\epsilon$-classical 
results 
after $30$ kicks, 
and different values of $\tau$ 
at fixed $k$ and  $\eta/\tau$. In some of the examined $\tau$ range, 
two different $\epsilon$-classical modes are simultaneously present, with 
different signs of the acceleration. In order to single out one of them,   
we have plotted the 	quantities $E_{\pm}$, which, in the classical case,  
are equal to the average energy after $30$ kicks, computed 
over those  rotors in the ensemble which, at the given time, have a positive 
(respectively, negative) momentum. 
In the quantum case, they are equal to the energy, computed 
in the state obtained by projecting the rotor state over the positive 
(respe., negative)
momenta. The main peak in the left-hand plot is due to a $(1,0)$ mode 
whose interval of existence and stability is, according to (\ref{stabcond}), 
$1.043<\tau/(2\pi)<1.261$. 
Both the classical and the quantum data sharply arise 
at the onset of the mode. The rise of the quantum data is actually milder, 
because the $\epsilon$-classical island has to grow beyond a size 
$\sim\epsilon$ in order to be quantally significant (note that 
$|\epsilon|$ increases on increasing $\tau$). As the stability border is 
approached, the island starts shrinking again, the classical and the 
quantum mode become less and less effective; the latter somewhat 
faster, for the just mentioned reasons. At the stability border, 
a $(2,0)$ classical mode is originated by bifurcation, which is 
in fact signalled by a tiny peak in the classical data; however, 
no similar quantum structure is observed.  
The small  peak on the 
leftmost part is an accelerator mode itself, see Sect. (3.2.6). 

At not really small values of the ``Planck's constant'' $\epsilon$, 
strong $\epsilon$-classical modes still leave quantal signatures. 
The quantum-classical correspondence turns however loose, in interesting 
ways. This is illustrated in the right hand part of Fig.\ref{fig6}. Now 
$| \epsilon |$ decreases from left to right. The rightmost peak corresponds 
to a $(3,-1)$ mode (see Sect. 3.2.6), faithfully reproduced by 
quantum data.  The main classical peak is the $(1,0)$ mode, active 
for $1.721<\tau/(2\pi)<1.862$. A significant quantum mode  
 is also observed, with remarkable  differences, however. In particular, 
the quantum mode has its maximum at the very  point where the $(1,0)$ 
$\epsilon$-classical mode turns unstable, giving birth to 
a $(2,0)$ mode, as demonstrated in Fig.2 (right).  
Our explanation of this curious behaviour is as follows.  
When the stability 
island near a stable fixed points breaks at a bifurcation point, some 
remnants of its KAM structure nevertheless survive, in the form of broken 
tori (cantori). These provide but a partial barrier to classical motion, 
and allow for nonzero phase-area flux. If this flux is small compared to 
$\epsilon$, then the cantorus quantally acts as if it were unbroken
\cite{cant} . Then it 
may give rise to a quantum accelerator mode, much more effectively than one 
might guess looking at the small area of stability 
of the bifurcated higher-period orbits.

Though $\epsilon-$classical modes exist at any value of $\beta$, their 
location
in the $\beta$-rotor's phase-space changes with $\beta$ . 
Hence their impact on the 
quantum evolution of a rotor state is enhanced at those values of 
$\beta$ which afford maximal overlap of the mode with the given state. 
In particular, 
 if $\eta\tau<2\pi$, and 
the $\beta-$rotor is initially set in the $n=0$ momentum 
eigenstate, 
then the quantum $(1,0)$-accelerator modes are   
especially  pronounced  when $\beta=l\pi/\tau-\eta/2$. 
Note that 
$\beta$ was not set to this value in the case of
of Fig.\ref{fig4}.  

\begin{figure}
\centerline{\epsfxsize=10cm\epsfysize=8cm\epsffile{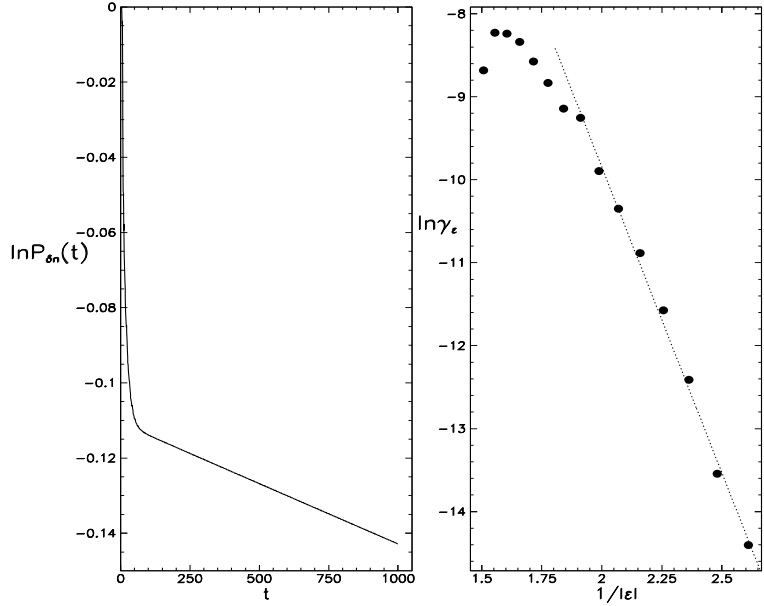}}
\caption{\small Exponential decay of accelerator modes. 
Left: semilogarithmic plot of 
the probability $P_{\delta n}(t)$ in a moving momentum window of width 
$\delta n=15$
centered at the  $\epsilon-$classical $(1,0)$ mode ,  versus time $t$, for 
$k=2.751$, $\tau=5.8$, $\epsilon=-0.483$, $\eta=0.092$, 
$\beta=\pi/\tau-\eta/2$.  
The asymptotic decay of $P_{\delta n}(t)$ is exponential with 
decay rate $\gamma_{\epsilon}\approx 0.32\times 10^{-4}$. 
Right: semilogarithmic plot of $\gamma_{\epsilon}$ 
vs $1/(|\epsilon|)$. 
The $\epsilon-$classical variables $\delta I=|\epsilon|\delta n$, $\kcl
=|\epsilon | k$ were held fixed at $7.540$, $1.329$ respectively while 
changing $\epsilon$. The fitting line corresponds to the law 
$\gamma_{\epsilon}\approx 135\times\exp(-7.35/|\epsilon|)$.
}
\label{fig7}
\end{figure}

In the $\epsilon-$ semiclassical regime, accelerator modes 
exponentially decay in time 
due to quantum tunneling out 
of the classical islands, with  decay rate   
$\gamma_{\epsilon}\propto \exp{-(\mbox{\rm const.}/|\epsilon|)}$. 
The decay of quantum accelerator modes  is numerically demonstrated in 
Fig.\ref{fig7}.  

\subsubsection{Quantum accelerator modes in Particle 
 Dynamics.}

Quantum Accelerator modes arise in the particle dynamics as well, just because 
such dynamics comes of a quantal superposition of $\beta-$rotors. 
We shall presently discuss  the small-$\epsilon$ 
asymptotics of the particle dynamics. This will at once provide  an 
alternative 
 derivation of the $\epsilon$-classical rotor 
dynamics, and a means of translating into particle dynamics  the results 
established in the previous sections.
   
Let us consider the propagator from state $|p\rangle$ to state $|x'\rangle$ 
from (discrete) time $t$ to time $t+1$, for the particle dynamics 
(\ref{fall7}). 
Let $p=n+\beta$ as usual, and 
$x'=2m'\pi+\theta'$ with $m'$ integer  and $0\leq\theta'<2\pi$.
 Then \begin{eqnarray}
\label{partprop}
\langle x'|\Uop(t)|p\rangle&=&\frac{1}{\sqrt{2\pi}}
 \exp(i\phi(\beta,\theta',t))\nonumber\\
&\times&\exp\left(-ik\cos(\theta')+in\theta'+2i m'\pi\beta
-\frac{i}{2}\epsilon n^2-i n\pi l-in\tau(\beta+t\eta+\frac{\eta}{2})\right)\;,
\end{eqnarray}
where $e^{i\pi ln^2}=e^{i\pi l n}$ was used, and
\begin{equation}
\label{phas}
\phi(\beta,\theta',t)=\beta\theta'-\frac{\tau}{2}
\left(\beta+t\eta+\frac{\eta}{2}\right)^2\,.
\end{equation}
Next we introduce $\epsilon$-classical scaled variables $I=n|\epsilon|$, 
$\kcl=k|\epsilon|$, $L'=-2\pi m' |\epsilon|$, to be kept constant in the 
$\epsilon$-classical limit. Then (\ref{partprop}) rewrites as:
\begin{equation}
\label{partprop1}
\frac{1}{\sqrt{2\pi}}\,
e^{ i\phi(\beta,\theta',t)}\,
e^{\frac{i}{|\epsilon|}{\mathcal F}(\theta',L',\beta,I,t)},
\end{equation}
where:
\begin{equation}
\label{genf}
{\mathcal F}(\theta',L',\beta,I,t)=I\theta'-L'\beta
-\kcl\cos(\theta')-\frac{1}{2}\mbox{\rm sign}(\epsilon)I^2
-I\left(\pi l+\tau(\beta+t\eta+\eta/2)\right)\,.
\end{equation}
Considering $|\epsilon|$ as the Planck's constant, and $I,L'$ as canonical 
momenta respectively conjugated to $\theta, \beta'$, the function 
${\mathcal F}$ 
is a generating function for the canonical transformation   
$(\theta,I,\beta,L)\to(\theta',I',\beta',L')$ given by: \begin{eqnarray}
\label{partmap}
\beta' \,&=&\,-\frac{\Dp{\mathcal F}}{\Dp L'}\,=\,\beta\nonumber\\
L\,&=&\,-\frac{\Dp{\mathcal F}}{\Dp \beta}\,=\,L'+\tau I\nonumber\\
I'\,&=&\,\frac{\Dp{\mathcal F}}{\Dp\theta'}\,=\,
I+\kcl\sin(\theta')\nonumber\\
\theta\,&=&\,\frac{\Dp{\mathcal F}}{\Dp I}\,=\,
\theta'-\mbox{\rm sign}(\epsilon)I-\pi l-\tau(\beta+t\eta+\eta/2)\;.
\end{eqnarray}
The second exponential in (\ref{partprop1}) is thus (apart from 
a constant prefactor) the $\epsilon$-semiclassical propagator associated 
with 
the $\epsilon$-classical map (\ref{partmap}). The 
3d and the 4th equation are just the $\epsilon$-classical $\beta$-rotor 
dynamics. The first equation says that quasi-momentum is conserved, 
and the second yields the complete revolutions accumulated  by the 
$\beta$-rotor from time $t$ to $t+1$; this quantity is formally conjugated 
to quasi-momentum. 

However, (\ref{partprop1}) has the additional 
phase factor $e^{i\phi}$, which is {\it not} $\epsilon$-semiclassical, 
because $\phi$ is not scaled by $|\epsilon|^{-1}$. 
While this factor is irrelevant 
for the  fixed $\beta$ dynamics (and was in fact disregarded in sect. (3.2.1)), 
it cannot be neglected when studying the particle wavepacket dynamics, 
which requires integration  over all 
values of $\beta$. Such integration causes 
 $\epsilon$-classical trajectories of 
different $\beta$-rotors   to 
interfere. This interference  is ruled by the true Planck's 
constant $\hbar=1$, 
and {\it cannot} be suppressed by the $\epsilon\to 0$ limit. Unlike the 
$\beta$-rotor dynamics, the particle dynamics does not become 
``classical'' in the $\epsilon\to 0$ limit. 

As long as $\beta$ is fixed, the maps (\ref{partmap}) yield, in the 
physical variables $p,x$ and at $|p|,|x|>>1$: 
\begin{eqnarray}
\label{partmap2}
p_t&=&|\epsilon|^{-1}I_t+\beta\sim p_0+\left(\frac{2\pi\jG}{\pG}\mp\tau\eta\right)
\frac{t}{|\epsilon|}\;,\nonumber\\
x_t&\sim& -|\epsilon|^{-1}L_t\sim x_0+
n_0\tau\, t+\left(\frac{2\pi\jG}{\pG}\mp\tau\eta\right)
\frac{\tau t(t-1)}{2|\epsilon|}
\end{eqnarray} 
for a $(\pG,\jG)$- accelerator mode (\ref{growth}) started at $x_0$, $p_0$.

\begin{figure}
\centerline{\epsfxsize=14cm\epsffile{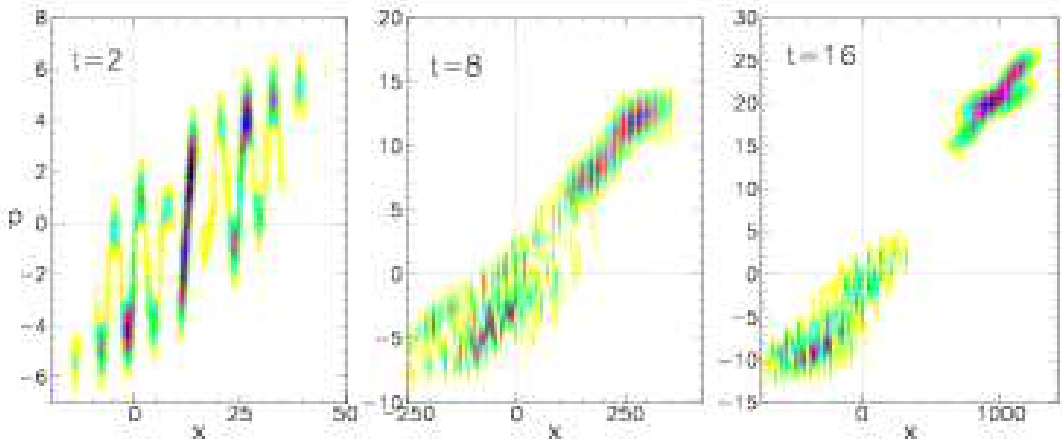}}
\caption{\small  Quantum  
phase space evolution for a particle initially prepared in the coherent 
state centered at $p_0 =0$, $x_0 =0.42$, with $k=\pi$, $\tau=5.86$,
 $\eta=0.093$. 
The computation implements the Bloch-Wannier fibration 
discretized over $512$ values 
of quasi-momentum. Shades-of-grey plots  of the Husimi function of the 
particle at  times $t=2,8,16$ are shown.
}
\label{fig8}
\end{figure}

Fig.\ref{fig8} shows the quantum 
evolution of a particle started at $t=0$ 
in the coherent state centered at the $(1,0)$-accelerator mode 
$p_0=0$, $x_0=0.42$. The 
 phase-space distribution splits, the righmost part of it 
moving with constant acceleration 
$-\tau\eta/\epsilon$. This is the effect of the accelerator mode.

A conceptually simpler situation is met when the initial state of the particle 
is an incoherent mixture of plane waves, for in that case 
no interference occurs between different $\beta-$rotors. Let the 
initial particle state  
be described in the falling frame  by the statistical operator:
\begin{equation}
\label{mix}
{\hat \rho}(0)\;=\;\int dp\;f(p)|p\rangle\langle p|\;,\;\;
f(p)\geq 0\;,\;\int dp\;f(p)=1\;.
\end{equation}
Each plane wave has a well-defined quasi-momentum, so it  is equivalent to a 
unique $\beta-$rotor in the angular momentum 
eigenstate specified by the integer part of the momentum of the wave.
Therefore, the statistical ensemble (\ref{mix}) is equivalent to a 
statistical ensemble of $\beta$-rotors. At any given quasi-momentum 
$\beta$, the state of the rotor is  described by the statistical 
operator ${\hat\rho}_{\beta}=(P(\beta))^{-1}\sum_n
 f(n+\beta)|n\rangle\langle n|$;  the distribution of quasi-momenta is 
further given by 
$P(\beta)=\sum_n f(n+\beta)$. 
The momentum distribution for the particle is given at time $t$ 
by 
\begin{equation}
\label{mixt}
f(p,t)=P(\beta)\langle n|{\hat\rho}_{\beta}(t)|n\rangle\;
\end{equation}
where $\beta=\{p\}$, $n=[p]$, and ${\hat\rho}_{\beta}(t)$ evolves according 
to the $\beta$-rotor dynamics (\ref{rotdyng}). The distribution in momentum 
is then $\epsilon$-quasi classically that of an ensemble of classical rotors 
evolving according to (\ref{clmap}).

\subsubsection{Spectroscopy of accelerator modes.}


\begin{figure}
\centerline{\epsfxsize=12cm\epsfysize=10cm\epsffile{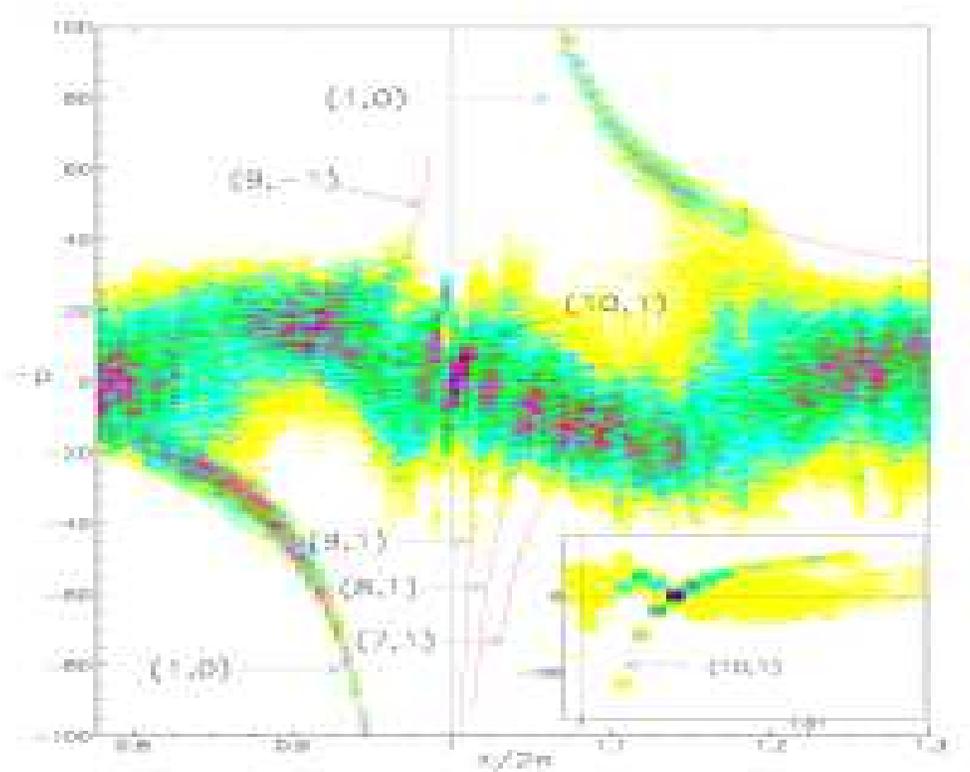}}
\caption{\small Momentum distribution in the falling frame 
after $60$ kicks for different values of the kicking period $\tau$ 
near $\tau=2\pi$, and for 
$k=0.8\pi$, $\eta=0.01579\tau$.
Note the negative sign of $p$. 
Darker regions correspond to higher 
probability. The initial state is  a mixture of $50$ plane waves sampled 
from a gaussian distribution of momenta. Full lines are the 
theoretical curves (\ref{curv}); 
they are drawn piecewise, to avoid hiding the actual structures to which 
they correspond. Their  order and  jumping index are 
indicated  by the arrows. The inset shows data at $t=400$.}
\label{fig9}
\end{figure}
\begin{figure}
\centerline{\epsfxsize=12cm\epsfysize=10cm\epsffile{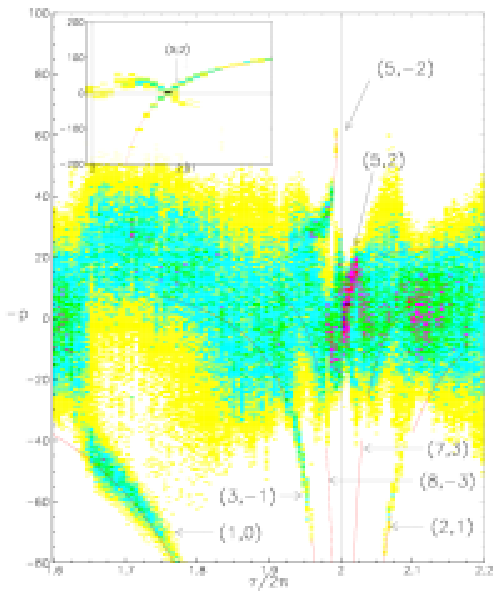}}
\caption{\small Same as Fig.\ref{fig9}, for $\tau$ near $4\pi$.
 The inset shows data at time $t=400$.}
\label{fig10}
\end{figure}

In the experiments described in refs.\cite{Ox1,Ox2,Ox3}, 
the initial state of the falling atoms is satisfactorily described, according 
to the same references, by (\ref{mix}), with $f(p)$ a Gaussian of 
rms deviation $\simeq 2.55$ (in our units) centered at $p=0$.  
 In Figs. \ref{fig9} and \ref{fig10} we show 
numerical results obtained with the same choice of the initial state. Such 
results provide further evidence 
of accelerator modes, including higher order ones, which were not 
previously  identified. The Figures were  produced by computing the evolution 
of an 
ensemble of $50$ rotors with the mentioned initial distribution over $60$ 
kicks, 
for different values of  the period $\tau$ near the main resonances 
$\tau=2\pi$, $\tau=4\pi$, and $k=0.8\pi$. As $\eta/\tau$ 
was kept fixed at the physical 
value $0.01579$, $\eta$ also varied with $\tau$. 
Fig.\ref{fig9} is analogous to  the experimentally obtained Fig.2
in ref. \cite{Ox1} (with some differences in units and in parameter values, 
though).

\noindent 
The final distribution 
of momenta is represented by shades of grey in the $(p,\tau)$ plane,   
darker zones corresponding to higher probability. 
The hyperbolic-like 
structures near $\tau=2\pi,4\pi$ are signatures of quantum accelerator 
modes. At any  $\tau$ where they are visible, they are in fact located 
at the momentum values reached at $t=60$ by certain 
accelerator modes,  started 
at $t=0$ with $n_0\simeq 0$. Such values are theoretically predicted by 
eqs.(\ref{growth}). The  
 $\epsilon$-classical $(\pG,\jG)$-mode started at $t=0$ with 
$I_0=n_0|\epsilon|$ is located at time $t$ at the momentum:
\begin{equation}
\label{curv}
n\simeq n_0- t\frac{\tau\eta}{\epsilon}+t\frac{2\pi\jG}{\pG |\epsilon|}\;.
\end{equation}
Replacing  $\epsilon=\tau-2\pi l$ ($l=1,2$ for Fig.\ref{fig9} and 
 Fig.\ref{fig10} respectively), and $\eta=0.01579\tau$, one obtains a   
curve $n=n(\tau)$  for any chosen 
time $t$ and for any chosen $n_0,\pG,\jG$. Such curve is 
then observable,  if a  mode with the chosen $\pG$,$\jG$ exists, 
which significantly overlaps the initial distribution. 
 The narrow distribution 
of initial momenta legitimates the choice $n_0=0$; although other small 
values of $n_0$ are involved, they just result in a thickening  of the 
curves. 

Well-marked (1,0) modes are observed in both Figures. 
The intervals of existence and stability of the $\epsilon$-classical 
$(1,0)$ modes predicted by (\ref{stabcond}) are $0.745<\tau/(2\pi)<0.963$ and 
$1.043<\tau/(2\pi)<1.261$ for the case of Fig.\ref{fig9}. 
The $(1,0)$ mode in Fig.9 is the same as in Fig.5; 
 in most of its  range 
the $\epsilon$-classical structure is like the one shown
 in Figs.\ref{fig2} and \ref{fig6}. 
It has both a  $(1,-1)$ stable 
orbit and a $(2,0)$ stable orbit originated by bifurcation. 
 The $(1,-1)$ 
theoretical 
curve mostly lies at negative values of $p$ off the scale  of the figure, 
and no significant trace of it was detected in our quantum 
computation, indicating that the island was too tiny compared to 
the relatively large values of $|\epsilon|$. 
We therefore explain the large $(1,0)$ structure obaserved by 
the same quantal mechanism discussed in sect. (3.2.4), namely 
localization by cantori.

Higher-order modes $(10,\pm1)$ and $(5,\pm2)$  
are observed close to $\tau=2\pi$ and $\tau=4\pi$ respectively. The 
corresponding $\epsilon$-classical modes are shown in Fig.\ref{fig3}.
 The small, yet well marked, structure 
produced by the $(10,1)$ mode near $2\pi$ 
is also visible in experimental data 
in \cite{Ox1}. The correspondence of the $(5,2)$- and $(10,1)$-modes  
 with the curves (\ref{curv}) is  remarkably evident 
at longer times, see the insets in Figs.\ref{fig9} and \ref{fig10}. 

In Figs.\ref{fig9} and \ref{fig10} other higher-order modes are visible, too.
 These were identified 
by first  fitting the observed structure with a curve (\ref{curv}), and then 
checking that the $\epsilon$-classical phase space really displays, 
in the relevant $\tau$ range, a  stable 
orbit with the found $\pG,\jG$. A few of these left but a dim trace 
in Figs.\ref{fig9} and \ref{fig10}. This may be due on one hand 
to mutual interference of different modes when they coexist in the same 
 $\tau$-range, and 
on the other hand to the small number of rotors used in the computation. 
For such reasons, Figs.\ref{fig9} and \ref{fig10}
 do not probably account for all the 
accelerator modes which are excited in their respective parameter ranges.

Generally speaking,  
the family of higher-order modes that are potentially observable in Figures 
like \ref{fig9} and \ref{fig10} (and in experiments as well)
 is probably much richer than shown here. 
These might be exposed by varying parameters, 
and also by a fine scan of 
 smaller $\epsilon$ ranges. It looks likely that momentum 
distributions at relatively short times, of the type shown 
in Figs.\ref{fig9} and \ref{fig10}, can be altogether
 described by {\it accelerator mode spectroscopy}, 
{\it i.e} identification of accelerator modes   and analysis of their 
mutual interference. Such a systematic analysis is beyond the scope 
of this work.

\section{Concluding remarks.}

\noindent 
1) The above theory of accelerator modes hinges on reduction to independent
kicked rotors models, wherein  such modes 
admit a natural interpretation in terms of classical 
trajectories. 
It is the conservation of quasi-momentum that allows for such reduction. 
Accelerator modes in particle dynamics are a purely quantal effect, in 
fact a remarkable manifestation of the conservation of quasi-momentum 
(in the falling frame).

\noindent
2) The fact that the intermediate-time dynamics is dominated by a discrete 
set of  
modes, which exponentially decay in time, bears a distinct resemblance 
to the Wannier-Stark problem of a Bloch particle in a constant field
\cite{WS}. How far this analogy carries is, in our opinion, 
a very interesting question.  

\noindent
3)  The time-dependent 
variants of the Kicked Rotor model examined  in this paper also raise 
other important theoretical questions, which were not addressed in this paper. 
These are about the  long-time asymptotics 
of the dynamics, and the localization-delocalization issue in particular.

\noindent
4) In the experiments \cite{Ox1,Ox2,Ox3} the initially prepared states 
form an {\em incoherent} superposition of  momentum eigenstates, and of 
quasimomentum states. Therefore the results of the of the calculations 
for the $\beta$-rotors explain the experiments. The particle nature 
(compared to the rotor) is 
important only for initial {\em coherent} superpositions of momentum
eigenstates. 
It will of great interest if this point is explored experimentally.

\section{Appendix: Resonances in the presence of Gravity}

In this appendix we assume $\tau=2\pi l$, with $l\neq 0$ integer.
We denote $\psi(x)=\langle x|\psi\rangle$ the wave function 
of the particle in the position representation, and $\En(t)$ the 
kinetic energy at time $t$ (in the falling frame). 
\par\vskip 0.3cm\noindent
{\bf Proposition:} {\it Let $\int dx\;|x|^{2\alpha}|\psi(x)|^2<+\infty$ 
for some $\alpha> 1$. Then:
\par\vskip 0.2cm\noindent
(I) If $\eta=0$ :
$$
\En(t)=\En(0)+\frac{k^2 Dt}{4}\; +\;O(t^{\lambda})\;,\;\;\;\;
\lambda=\Mx\{2-\alpha,\frac{1}{2}\}  
$$
where: 
\begin{equation}
\label{prop1}
D\;=\;\frac{1}{l}\sum\limits_{n=0}^{l-1}\int_0^{2\pi}d\theta
|\langle\theta|\Psi_{\beta_n}\rangle|^2\;\;,\;\;\beta_n=\frac{n}{l}+\frac{l}{2}\;
\mbox{\rm mod}(1)\;.
\end{equation}
(II) If $\eta\neq 0$ satisfies a Diophantine condition, that is,  
there are constants 
$c,\gamma>0$ so that for all integer $p,q$ :
\begin{equation}
\label{dio}
\left\vert \eta-\frac{p}{q}\right\vert\;\geq\;c\;q^{-2-\gamma}\;
\end{equation}
then:
\begin{equation}
\label{prop2}
\En(t)=\En(0)+\frac{k^2t}{4}+O(t^{\sigma})\;,\;\;\;\;\sigma=\Mx\{2-\alpha+\gamma,\frac{1}{2})
\end{equation}
}

\par\vskip 0.2cm\noindent
{\bf Remarks:} 

\noindent 1.  The quasi-momenta $\beta_n$'s 
in part (I) are exactly those yielding quadratic growth of the $\beta-$rotor 
energy.

\noindent 2. Part (II) ensures that, for all 
$\eta$ in a set of full measure,  
the energy grows diffusively with coefficient $k^2/4$.
\par\vskip 0.2cm\noindent
{\it Proof:} Numerical constants  
will share the common 
notation $C$ whenever their exact value is irrelevant. 
 From (\ref{rotopg}), 
$$
\langle n|\Rp_{\beta}(t)|\Psi_{\beta}(t)\rangle\;=\;
e^{-i\pi l n(1+2\beta+2t\eta+\eta)}e^{i\phi(\beta,\eta,t)}\langle n|\Psi_{\beta}
(t)\rangle\;
$$
The explicit form of the phase $\phi$ is not important for our present 
purposes. Further, 
$$
\langle\theta|\Psi_{\beta}(t+1)\rangle\; 
=\langle\theta|\Kp\Rp_{\beta}(t)|\Psi_{\beta}(t)\rangle\;=\;
e^{i\phi(\beta,\eta,t)}e^{-ik\cos(\theta)}\langle \theta-a-bt |\Psi_{\beta}(t)\rangle\;,
$$
where:
\begin{equation}
\label{ab}
a\;=\;\pi l(1+2\beta+\eta)\;\;,\;\;b=2\pi l\eta\;.
\end{equation}
 It follows that:
$$
\langle\theta |\Psi_{\beta}(t)\rangle\;=\;
e^{-i\phi(\beta,\eta,t)}e^{-ik F(\theta,\beta,t)}\langle\theta-at-\frac {b}{2}t(t-1)|\Psi_{\beta}(0)\rangle
$$
where:
$$
F(\theta,\beta,t)\;=\;\sum\limits_{r=0}^{t-1}\cos (\theta-ra-rbt+r^2b/2+rb/2) 
$$
We now use eqn.(\ref{mess}). As already remarked, the dominant contribution 
is given 
by the 1st term on the rhs, corrections being on the order of square root 
of that term. 
 We hence restrict to that term. After 
substituting the above equations, it takes the form: 
\begin{equation}
\label{etot}
\frac{1}{2}\int_0^1d\beta\int_0^{2\pi}d\theta\;
\left|\frac{d}{d\theta}\langle\theta|\Psi_{\beta}(t)\rangle\right|^2
=\En_0(t)+\En_1(t)+\En_2(t),
\end{equation}
having denoted
$$
\begin{array}{ccc}
\En_0(t)&=&\frac{1}{2}
\int_0^1d\beta\int_0^{2\pi} d\theta |g'(\theta,t)|^2\;,\\
\En_1(t)&=&\frac{k^2}{2}\int_0^1d\beta\int_0^{2\pi} d\theta
F^{'2} (\theta ,\beta,t)|g(\theta,t)|^2\;,\\
\En_2 (t)&=&k\;\Re\int_0^1d\beta\int_0^{2\pi} d\theta\;
iF'(\theta,\beta,t)g^*(\theta,t)g'(\theta,t)\;\\
g(\theta,t) &=&\langle\theta-ta-bt^2/2+bt/2|\Psi_{\beta}(0)\rangle 
\end{array}
$$
(primes denote differentiation  with respect to $\theta$). An obvious shift 
in 
$\theta$ shows that 
\begin{equation}
\label{ezero}
\En_0(t)=\mbox{\rm const}=\En_0(0). 
\end{equation}
Moreover, 
from the Cauchy-Schwarz
inequality it 
follows that:
\begin{eqnarray}
\label{euno}
|\En_2(t)| &\leq& k\left(\int_0^1d\beta \int_0^{2\pi} d\theta
F^{'2} (\theta,\beta,t)|g(\theta,t)|^2\right)^{1/2}
\left(\int_0^1d\beta\int_0^{2\pi} d\theta 
|g'(\theta,t)|^2\right)^{1/2}\nonumber\\
&=& 2\En_1(t)^{1/2}\;\En_0(0)^{1/2}
\end{eqnarray}
The dominant contribution to $\En(t)$ is thus given by $\En_1(t)$, so 
$\En_0$, $\En_2$ will not be  considered until the end of the proof. 
After an obvious change of variables,
\begin{equation}
\label{e1}
\En_1(t)\;=\;\frac{k^2}{2}\int_0^1d\beta\int_0^{2\pi} d\theta
\left(\sum\limits_{r=1}^{t}\sin(\theta+ra+br^2/2-br/2)\right)^2|\langle\theta|
\Psi_{\beta}(0)\rangle|^2\;.
\end{equation}
We rewrite the square of the sum  
as a double sum, then apply standard trigonometric formulae, and finally replace 
$a,b$ by (\ref{ab}), leading to :
\begin{equation}
\label{e11}
\En_1(t)\;=\;\frac{k^2}{2}\int_0^1 d\beta\int_0^{2\pi}d\theta \;
|\langle\theta|\Psi_{\beta}(0)\rangle|^2\;
\left(A_1(\beta,t)-A_2(\beta,\theta,t)
\right)
\end{equation}
with
$$
\begin{array}{ccc}
\label{dbsum}
A_1(\beta,t)&=&\frac{1}{2}\sum\limits_{r,s=1}^{t}
(-)^{(r+s)l}\cos\left(2\pi l\beta(r-s)+\pi l\eta(r^2-s^2)\right)\;,\nonumber\\
A_2(\beta,\theta,t)&=&
\frac{1}{2}\sum\limits_{r,s=1}^{t}(-)^{(r+s)l}
\cos\left(2\theta+2\pi\beta(r+s)+\pi l\eta(r^2+s^2)\right)\;.
\end{array} 
$$
Now we expand $|\langle\theta|\Psi_{\beta}(0)
\rangle|^2$ in Fourier series:
\begin{equation}
\label{fs}
|\langle\theta|\Psi_{\beta}(0)
\rangle|^2\;=\;\sum\limits_{M,N}\;c(M,N)\;e^{2\pi iM\beta}\;e^{iN\theta}\;
\end{equation}
Replacing  in (\ref{e11}) we obtain  
\begin{equation}
\label{esti}
\En_1(t)\;=\;\frac12\pi k^2\Re(B_1-B_2)\;,
\end{equation}
where:
\begin{eqnarray}
\label{B1}
B_1\;&=&\;\sum\limits_{r,s=1}^{t}
c(l(r-s),0)(-)^{(r-s)l}e^{-i\pi l\eta(r^2-s^2)}\\
&=&\;\sum\limits_{j=1-t}^{t-1}c(lj,0)(-)^{lj}e^{i\pi l\eta j^2}
\sum\limits_{r=\Mx(1,j+1)}^{\Mi(t,j+t)}e^{-2\pi il\eta rj}\;,\nonumber\\
B_2\;&=&\;\sum\limits_{r,s=1}^{t}c(l(r+s),2)(-)^{l(r+s)}e^{-i\pi l\eta(r^2+
s^2)}\nonumber\\
&=&\;\sum\limits_{j=2}^{2t}c(lj,2)(-)^{lj}e^{-i\pi l\eta j^2}
\sum\limits_{r=\Mx(1,j-t)}^{\Mi(t,j-1)}e^{-2\pi i l\eta r(r-j)}\;.
\end{eqnarray}
With the help of the Lemma proven in the end of this section, $B_2$  is 
bounded by :
$$
\sum\limits_{j=2}^{1+t}|c(lj,2)|(j-1)+\sum\limits_{j=t+2}^{2t}
|c(lj,2)|(2t-j+1)\leq C\sum\limits_{j=2}^{2t}(j-1)j^{-\alpha}+
Ct\sum\limits_{j=t+2}^{2t}j^{-\alpha}\;=\;O(t^{2-\alpha})\;.
$$
Here and in the following $O(t^x)$ has to be read as $O(\log t)$, 
$O(1)$ whenever $x=0$, $x<0$ respectively. 
In order to estimate $B_1$ we distinguish two cases;
\par\vskip 0.2cm\noindent
{\it Case I: $\eta=0$.}
\begin{eqnarray}
\label{Bsum1}
B_1\;=\;\sum\limits_{j=1-t}^{t-1}c(lj,0)(-)^{lj}(t-|j|)
&=& t\sum\limits_{j=-\infty}^{\infty}(-)^{lj}c(lj,0)+O(t^{2-\alpha})
\nonumber\\
&=& \frac{Dt}{2\pi}+O(t^{2-\alpha})
\end{eqnarray}
where 
\begin{eqnarray}
\label{Dsum}
D\;&=&\;2\pi \sum\limits_{j=-\infty}^{\infty}
(-)^{lj}c(lj,0)\nonumber\\
&=&\;\frac {2 \pi}{l}\sum\limits_{n=0}^{l-1}\sum\limits_{j'=-\infty}
^{\infty}c(j',0)e^{\pi ij'(2n/l+l)}\;\nonumber\\
&=&\;
\frac {1}{ l}\sum\limits_{n=0}^{l-1}\int_0^{2\pi} d \theta
|\langle\theta|\Psi_{\beta_n}\rangle|^2\;\;,\;\;\beta_n=\frac{n}{l}+\frac l2\;
\mbox{\rm mod}\;(1)\;.
\end{eqnarray}
We next substitute (\ref{Bsum1}) and (\ref{Dsum}) in (\ref{esti}), and then 
in (\ref{etot}). Recalling  (\ref{ezero}),(\ref{euno}), and the remark 
preceding (\ref{etot}) leads to (\ref{prop1}).
\par\vskip 0.2cm
\noindent
{\it Case II: $\eta$ a Diophantine irrational}. 
The sum (\ref{B1}) is written in the form 
$$
B_1\;=\; c(0,0)t+S
$$
where $S$ is the contribution of
all $j\neq 0$ terms. It can be bounded as 
\begin{eqnarray}
|S|\;&\leq&\;2\sum\limits_{j=1}^{t-1}|c(jl,0)|\frac{2}{|\sin(\pi l\eta j)|}
\nonumber\\
&\leq&\;C\sum\limits_{j=1}^{t-1}|c(jl,0)|j^{1+\gamma}\;=\;O(t^{2-\alpha 
+\gamma}).
\end{eqnarray}
where (\ref{dio}) was used. The normalization of the wavefunction
implies $c(0,0)=1/(2\pi)$. 
Substituting in (\ref{esti})  
leads to  (\ref{prop2}) after 
the same concluding  steps  as in Case I above.  
This completes the proof. $\Box$ 
\par\vskip 0.3cm\noindent
{\bf Lemma.} {\it Under the hypotheses of the Proposition, and with 
$c(M,N)$ defined as in (\ref{fs}), $|c(M,N)|=O(|M|^{-\alpha})$ as 
$M\to\infty$.} 
\par\vskip 0.2cm\noindent
{\it Proof.} From the Bloch-Wannier fibration (\ref{psibeta}) it follows that, if
$|M|\geq 1$, 
then: 
\begin{eqnarray}
|c(M,N)|&=& \frac{1}{2\pi}\left\vert\int dx\;\psi^*(x+M\pi)\psi
(x-M\pi)\;e^{-iNx}\right\vert\nonumber\\
&\leq&
C |M|^{-\alpha}\int dx\;
|\psi^*(x+M\pi)\psi(x-M\pi)|Q(x,M)\;
\end{eqnarray}
where $Q(x,M)=(1+(x+M\pi)^2)^{\alpha/2}(1+(x-M\pi)^2)^{\alpha/2}\geq
(4\pi^2M^2)^{\alpha/2}$. 
The Cauchy-Schwarz inequality then yields :
$$
|c(M,N)|\leq C|M|^{-\alpha}
\int dx\;|\psi(x)|^2\;(1+x^2)^{\alpha}<\infty\;,
$$
as the convergence of the integral was  assumed in the
Proposition. $\Box$                     
\par\vskip 0.5cm\noindent
\noindent{\bf Acknowledgments.} 
This research was supported in part by 
{\it PRIN-2000: Chaos and localisation in classical and quantum mechanics}, 
by the US-Israel Binational Science 
Foundation
(BSF), by the US National Science Foundation under Grant No. PHY99-07949, 
by the
Minerva Center of Nonlinear Physics of Complex Systems, 
by the Max Planck Institute for
the Physics of Complex Systems in Dresden, 
and by the fund for Promotion of Research
at the Technion. Useful discussions with 
M. Raizen, M. Oberhaler and Y. Gefen are acknowledged.

\par\vfill\eject


\begin{thebibliography}{IzShep}


\bibitem{KR} for reviews see, e.g.: F.M. Izrailev, Phys. Rep. {\bf 196}, 299
(1991); S. Fishman, in {\it Proceedings of the International School of Physics 
Enrico Fermi: Varenna Course CXIX}, G.Casati, I. Guarneri and U.Smilansky 
eds., North Holland 1993, p.187.

\bibitem{chirikov} B.V.Chirikov, Phys.Rep. {\bf 52}, 263 (1979).

\bibitem{lichtenbergb} 
A.J. Lichtenterg and M.A. Liberman, {\em Regular and Chaotic Dynamics}, (Springer-Verlag, NY, 1992)


\bibitem{ZE} G.M. Zaslavsky, M. Edelman, and B.A. Niyazov, 
Chaos {\bf 7}, 159 (1997); G. M. Zaslavsky and M. Edelman, 
Chaos {\bf 10}, 135 (2000). 
\bibitem{FGP} S.
 Fishman, D.R. Grempel, and R.E. Prange, Phys. Rev. Lett.
{\bf 49}, 509 (1982); D.R. Grempel, R.E. Prange, and S. Fishman, Phys.
Rev. A {\bf 29}, 1639 (1984).

\bibitem{hoa} J.D. Hanson, E. Ott, and T.M. Antonsen,
Phys. Rev. A {\bf 29}, 819 (1984); A. Iomin, S. Fishman and G. Zaslavsky,
to be published in Phys. Rev. {\bf E}. 


\bibitem{IzShep} F.M.Izrailev and D.L.Shepelyansky, Sov. Phys. Dokl. 
{\bf 24}, 996 (1979); G.Casati and I.Guarneri, Comm. Math. Phys. 
 {\bf 95}, 121 (1984). 


\bibitem{IEEE} G.Casati, B.V. Chirikov, D.L. Shepelyansky, and 
I.Guarneri, Phys. Rep. {\bf 154}, 2 (1987); 
G.Casati, I.Guarneri and D.L.Shepelyansky, 
IEEE J. Quantum. Electron. {\bf 24}, 1420 (1988), and references therein.

\bibitem{expH} E.J.Galvez, J.E.Sauer, L.Moorman, P.M..Koch, and D.Richards, 
Phys. Rev. Lett. {\bf 61}, 2011 (1988); 
J.E.Bayfield, G.Casati, I.Guarneri, and D.W.Sokol, Phys. Rev. Lett.
{\bf 63}, 364 (1989); M.Arndt, A.Buchleitner, R.N.Mantegna, and 
H.Walther, Phys. Rev. Lett. {\bf 67}, 2435 (1991). 
 
\bibitem{raizen1}
D.A. Steck, V. Milner, W.H. Oskay, and M.G. Raizen,
Phys. Rev. {\bf E 62}, 3461 (2000);
F.~L. Moore, J.~C. Robinson, C.~F. Bharucha, Bala Sundaram, and
M.~G. Raizen, Phys. Rev. Lett. {\bf 75}, 4598 (1995);
C.~F. Bharucha, J.~C. Robinson, F.~L. Moore, Qian Niu, Bala Sundaram, and
M.~G. Raizen, Phys. Rev. {\bf E 60}, 3881 (1999);
B.G. Klappauf, W.H. Oskay, D.A. Steck amd M.G. Raizen, Physica 
(Amsterdam) {\bf 131 D}, 78 (1999).

\bibitem{GSZ}
R. Graham, M. Schlautmann and P. Zoller, Phys. Rev. {\bf A 45}, R19 (1992).

\bibitem{CT}, C.Cohen-Tannoudji and J.Dupont-Roc, 
{\it Atom-Photon interactions: basic processes and applications},
 Gilbert Grynberg 1992. 


\bibitem{BFS} R. Blumel, S. Fishman and U. Smilansky, 
J. Chem. Phys. {\bf 84}, 2604-2614 (1986).


\bibitem{phirot} F.M. Izrailev, Phys. Rev. Lett. {\bf 56}, 541 (1986).

\bibitem{raizen2}
W.~H. Oskay, D.~A. Steck, V. Milner, B.~G. Klappauf, and M.~G. Raizen,
Opt. Comm. {\bf 179}, 137 (2000).



\bibitem{WGF} S.Wimberger, I.Guarneri and S.Fishman, in preparation.

\bibitem{Ox1} M.K. Oberthaler, R.M.Godun, M.B. d'Arcy, G.S. Summy, 
and K. Burnett, Phys. Rev. Lett. {\bf 83}, 4447 (1999).

\bibitem{Ox2} R.M.Godun, M.B. d'Arcy, M.K. Oberthaler, G.S. Summy, 
and K. Burnett, Phys. Rev. A{\bf 62}, 013411 (2000).


\bibitem{Ox3}  M.B. d'Arcy, R.M.Godun, M.K. Oberthaler, G.S. Summy,
and K. Burnett, S.A. Gardiner, Phys. Rev. {\bf E 64}, 056233 (2001)

\bibitem{qpp} D.L. Shepelyansky, Physica D{\bf 8} 208 (1983); 
G. Casati, G.Mantica and D.L.Shepelyansky, Phys. Rev. E{\bf 63} 
066217 (2001), and references therein.



\bibitem{LL} L.D. Landau and E.M. Lifshiz, {\it Quantum Mechanics}, 3d 
edition 
(Pergamon, Oxford 1977), p.76.

\bibitem{cant} T. Geisel, G.Radons and J.Rubner, Phys. Rev. Lett. 
{\bf 57}, 2883 (1986); 
R.S. MacKay and J.D. Meiss, Phys. Rev. A{\bf 37}, 4702
( 1988); J.D. Meiss, Phys. Rev. Lett. {\bf 62}, 1576 (1989);
 D.R. Grempel, 
S. Fishman and R.E. Prange, 
Phys. Rev. Lett. {\bf 53}, 1212 (1984);
S. Fishman, D.R. Grempel and R.E. Prange, 
Phys. Rev. {\bf A 36}, 289 (1987).




\bibitem{WS} for a review see, e.g., G. Nenciu, 
Rev. Mod. Phys. {\bf 63}, 91 (1993) and references therein. 

\item S. Fishman, D.R. Grempel and R.E. Prange, Phys. Rev. {\bf A36}, 289-305 (1987).



\end{thebibliography}
\end{document}